# YBCO Transition Edge Sensors Used For Very Low Noise Thermal Control

B. Guillet, D. Robbes, L. Méchin

*Abstract*—In electrically substituted radiometers, the ultra low noise control of the temperature is required. In this framework, we present results dedicated to the temperature regulation of copper plates, 3 cm diameter and 1 mm thick, using $YBa_2Cu_3O_{7-\delta}$ (YBCO) transition edge sensors (TES). One of the TES was used as the active thermometer of the temperature control unit, whereas the two others measured the plate temperature behavior. Two heating resistors were wound along a spiral pattern, just underneath the copper plate, to minimize the heating delay. The correlation between the two TES was clearly highlighted by applying a small heat perturbation through the second distributed resistor, the path of which closely follows that of the main resistor. Calibrated temperature oscillations of 30 µK rms @ 10mHz together with spectral analysis were measured and a temperature resolution in the range of a few µK rms was achieved.

*Index Terms*—Radiometry, Thermal control, Thermistors

## I. Introduction

MANY sensitive instruments have their low frequency performances improved provided that some of their key parts or sub-systems benefit of low noise temperature control. It is particularly true in the field of metrological instruments. A good example is that of absorbing cavities used in absolute radiometers where the temperature fluctuations are a strong limiting factor [1], [2]. The same limiting factor also arises in more popular applications of microbolometers such that of infrared imaging systems [3]. On another hand, progress towards wide band absolute thermometers, with mK accuracy, also needs very efficient but convenient control of the sample holder. A temperature resolution well below the mK has to be reached in that case.

We present here a work dedicated to the temperature control of copper plates, 3 cm diameter and 1 mm thick, which is a first step towards an electrical substitution radiometer (ESR) in our laboratory. The paper is arranged as follows: section II gives an analysis of the rules that have led to the design of the sample holder. The third section gives an overview of electronics, both the read-out and the control units. Section IV is devoted to the results we have got using this assembly, fully tested YBCO transition edge sensors.

## II. Sample Holder

*A. Design*

Many parameters are of importance to design a temperature controlled system, even in the case of small copper or sapphire pieces coupled to small cryocooler heads receiving cryogenic chip mounts for various sensing purposes. The list of the main parameters is reported in Table I. It is adequate for quasi first order thermal systems. The numerical values of Table I are those of our system. Numerous links exist between these parameters, some of which must be specified at the beginning of the design. The heating power $P_H$ must be consistent with the evacuated power of the cooling system. $P_H$ is also related to the output voltage applied to the heating resistor $R_{H1}$. Since the typical values of $R_{H1}$ range from 10 to 50 Ω, it follows that, for a low noise operation, $P_H$ may reach a few tens of Watts. A low noise operation at this level means that standard low noise analog IC circuits and components must be used, ±12 V battery powered, with maximum output currents lower than a few Amps. From the maximum available heating power and the needed value for the maximum temperature variation $\Delta T_{max}$, the thermal conductance $G_t$ is deduced as their ratio. After these first parameters are fixed, the thermal capacitance $C_t$ and the time constant $\tau$ can be evaluated, provided that additional restrictions are introduced. Our sample holder is designed for an operating temperature above the liquid nitrogen temperature (77 K). At this elementary level of the design, the variations of some parameters of the sample holder model with the temperature are not introduced.

B. Guillet, D. Robbes and L. Méchin are with the Institut des Sciences de la Matière et du Rayonnement, GREYC UMR 6072, Caen, F-14050 FRANCE ( 33-2-31-45-26-95; fax: 33-2-31-45-26-98; e-mail: bguillet@ greyc.ismra.fr).

TABLE I
SAMPLE HOLDER THERMAL CHARACTERISTICS

| Symbol | Designation | Numerical value |
|---|---|---|
| $C_t$ | Thermal capacitance | 7.7 J/K |
| $G_t$ | Thermal link between $C_t$ and heat sink | 77 mW/K |
| $\tau$ | Time constant | 100 s |
| $T_0$ | Temperature of the heat sink | 77.4 K |
| $T$ | Temperature of the sample holder | 80 – 130 K |
| $\Delta T$ | Temperature variation | 10 – 20 K |
| $\Delta T_{max}$ | Maximal temperature variation | ~ 50 K |
| $R_{H1,2}$ | Heater electrical resistance | 30 Ω |
| $P_H$ | Heating power | Up to a few Watts |
| $\tau_d$ | Time delay | 0.4 s |

Moreover, the sample holder mass must be higher than the small system it will support (the cavity of a radiometer for example) in order to act as a heat sink. One point is extremely important, for the minimization of the temperature fluctuations around the working temperature value using a servo-loop: it is related to the inherent time delay $\tau_d$ between the heat production inside $R_H$ and the measurement of the related temperature variations by the thermometer in the loop. From the basis of servo loop theory, it is well known that the larger the open loop gain is, the better the rejection of heat perturbation is, provided that the Nyquist criterion is respected with a sufficient margin. For a first order system, this gain becomes finite as soon as a finite delay exists and the maximal gain value is related to the separation, in time domain, between the time constant and the delay. Furthermore, the time constant cannot be made infinitely large, and the design must be very careful with respect to the minimisation of the delay. As usual, pure delays are associated to propagation phenomena's, which is heat diffusion in our case. After choosing a high diffusivity $D$ material, minimising $\tau_d$ implies the distribution of the heat production just under the usable area of the sample holder. Indeed, $\tau_d$ scales with the ratio $l^2/D$ where $l$ is the length separating the front of the sample holder from the outset of the heat production. However, it cannot be made arbitrary short: it must be stiff enough to secure mechanical rigidity.

We also included two sub-systems designated to a proper identification and characterisation of the low noise system. It represents an important part of the system design. The first part is a second heating resistor $R_{H2}$, that is very well balanced and intertwined with the first one $R_{H1}$. It is used to introduced very well calibrated heat perturbations through the thermal system, both in open or closed configurations. The design is such that the Joule heatings through $R_{H1}$ or $R_{H2}$ are nearly identical. This allows a very convenient way to identify the system, because sinusoidal heat perturbations at twice the applied voltage frequency across $R_{H2}$ may be applied, due to the squaring from the Joule's Law. The second subsystem, very efficient for a proper characterisation of the temperature, consists of a pair of two auxiliary independent thermometers, having a very high sensitivity and monitored simultaneously. The coherence between the records lead to an improved system identification of the temperature fluctuations of the sample holder. Finally, the ratio between the amplitude of the detected temperature variations that result from the heat perturbation in the open and closed loop configurations gives a proper characterisation of the servo-loop efficiency.

*B. Fabrication*

The sample holder we built using the previous rules is schematically drawn in Fig. 1. The associated parameter values are listed in Table I. The usable area has a diameter of about 3 cm, and a ring made using standard printed circuit board allows easy electrical connections. The heating resistor $R_{H1}$ and $R_{H2}$ were made using constantan wires, intertwined, wound and glued in a spiral digged at the rear of the copper plate. The thermal insulation from the copper head of the cooler is made using an other disk of printed circuit, 1.27 mm thick, glued to the previous copper plate and a final one, clamped to the cold finger. All the system is in vacuum, surrounded by a 77 K radiation shield. The thermometers are three high temperature coefficient (TCR) resistors made using 200 nm thick high quality $YBa_2Cu_3O_{7-\delta}$ film deposited on a strontium titanate substrate by pulsed laser deposition. They are patterned in strips (600 µm long and 40 µm width) with four Au contacts. The total area of each thermometer is 1 mm$^2$. Alternatively, a industrial standard platinum resistor IPRT is mounted.

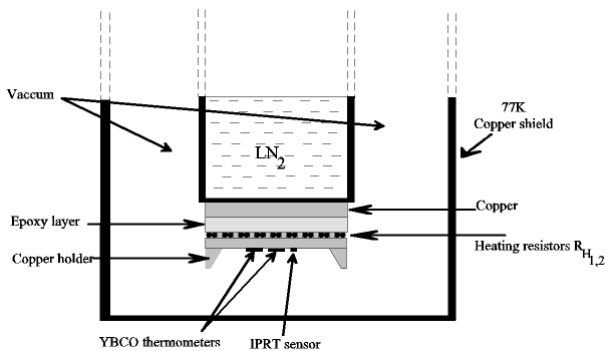

Fig. 1. Cut view of the sample holder. The important sub-systems are the use of two identical intertwined heating resistors, and the use of two independent, out of loop, thermometers.

## III. ELECTRONICS

The read-out electronics was developed in our laboratory, especially designed for 10 – 100 Ω thermistors, biased using highly stable square wave currents at 1 kHz and 1 mA amplitude. It has been fully tested at 300 K, showing that our system exhibits a temperature resolution of 25 $\mu K_{rms}$ in a bandwidth of 1 Hz and over a 2000 s measuring time. These results were obtained using IPRT sensor. Tested using resistors having a very low TCR, the electronic noise floor has been characterised in the time and the frequency domains. Made using standard low noise circuits, the voltage (white) noise, referred at the input, is 2 nV/√Hz, little dependent on the Johnson noise of the thermistor (< 100 Ω at 300 K). The excess low frequency noise is ranging between 1.8/√f (nV/√Hz) and 3.2/√f (nV/√Hz). Dividing the previous voltage noise spectrum by the thermistance responsivity (product $RI_b$xTCR) leads to the noise equivalent temperature (NET) of the system for an ideal (i.e. without excess noise) thermistor $R$, working under a bias current $I_b$. If we introduce the TCR of $YBa_2Cu_3O_{7-\delta}$ TES sensors that exhibit values of the order of 1 $K^{-1}$ and up to 4 $K^{-1}$ in the literature [4], one finds limiting NET values associated to our electronic read-out, lying between 5 to 20 nK, at 1 mA bias and for frequencies above a few Hz.

From the measured basic parameters $G_T$, $C_T$ and delay $\tau_d$, a PSPICE [5] model of the overall system was built, including a standard analog PID controller to close the servo-loop. The parameters of the PID were then calculated in order to match the conditions discussed in the previous section, and the final circuit was designed using standard methods in electronic design.

## IV. RESULTS

As pointed out in section two, a strategy is done to conveniently demonstrate the low noise properties of the sample holder.

It is associated to the use of the well balanced heating resistor $R_{H2}$ and the use of two independent, out of loop, thermometers together with the in loop one. Fig. 2 is a plot of the $R$ versus $T$ characteristics of the YBCO thermometers together with their derivative ; it shows very little dispersion, compatible with the experiments to be done with a high sensitivity along the superconducting transition. A TCR of about 1 $K^{-1}$ is easily deduced from the Fig. 2 plots.

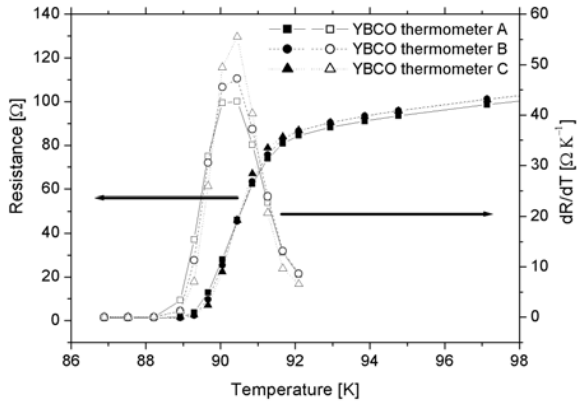

Fig. 2. Superconducting transition of the three $YBa_2Cu_3O_{7-\delta}$ strips used in the experiments at 1 mA bias (filled symbols). The strips are 600 μm long, 40 μm wide and 200 nm thick on a $SrTiO_3$ substrate. Also shown is the derivative with respect to temperature (closed symbols).

The plots of Fig. 3 show the effect of a sinusoidal excitation in the auxiliary resistor $R_{H2}$ that induces a heat perturbation in the thermal circuit at twice the frequency of the applied perturbing voltage. The frequency doubling appears very clearly, together with the very similar responses of the two independent thermometers. From such sets of measurements in closed loop, we deduced the efficiency of the heat perturbation rejection to be 400 at dc, in a flat bandwidth of about 50 mHz. Our system making good use of liquid nitrogen at ambient pressure as the coolant, we know that any pressure drift over the liquid nitrogen bath will be reflected as a temperature drift at the cold head, due to the Clapeyron's law. The conversion rate is 83 μK/Pa [6] and the atmospheric pressure drift and fluctuations may reach 1 Pa/s on windy days. This means that we are expecting temperature drifts as high as 80 μK/s on open loop and about 200 nK/s using the servo mode. The plots of Fig. 4 clearly show a temperature drift of 100 nK/s during the 900 s measuring time. This value is within a factor 2 below the expected one, for windy days, but at this level, we had not monitored the pressure to accurately check this prediction. This indicate that a pressure control above the liquid nitrogen bath would probably improve the very low frequency system performances. Superimposed to the low frequency drift in Fig. 4 is a temperature oscillation, 30 µK$_p$, at a frequency of 10 mHz. It is associated to a small heat perturbation at a frequency of 5 mHz in the resistor $R_{H2}$. These plots clearly show the ability of the thermometer pair to follow very small and very low temperature variations. Suppressing the heat excitation shows the temperature drift and small differences between the observing thermometers. It is plotted in Fig. 5. Again the rate of the temperature drift is of the order of 100 nK/s and some fluctuations around the mean drift value appear still correlated. They are in the range of 10-20 µK$_{pp}$. The difference between the two thermometer indications has also been plotted, it shows a maximum value of ± 6 µK, with a standard deviation of 2.8 µK$_{rms}$ calculated in this 400 s measuring time. These values are much higher than those expected from the known noise of the detecting electronic (see section III). The clear identification of the origin of both the coherent part of the fluctuations around the mean drift and the non coherent part observed in the difference is still under investigation. We nevertheless have preliminary elements that indicate that the coherent deviation around the mean drift reflects true temperature variations of the liquid nitrogen bath, efficiently rejected by the servo-loop. Some other elements indicate that the non coherent fluctuations around the mean drift appear rather associated to a large excess noise at very low frequency in the YBa$_2$Cu$_3$O$_{7-\delta}$ film operated at the resistive transition.

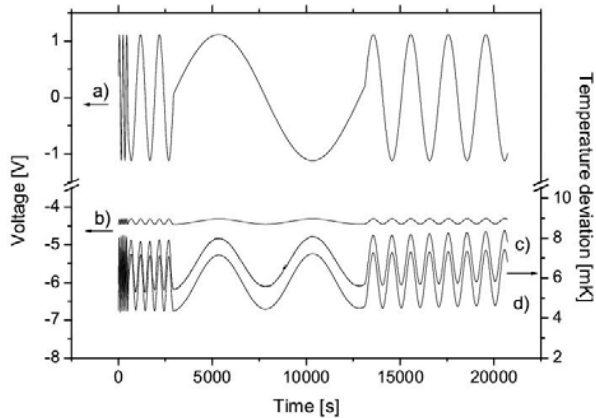

Fig. 3. Curve (a): sine wave voltage applied to the heating resistor $R_{H2}$ Curve (b) is the in loop thermometer response and curve (c, d), referred to the right axis, are the out of loop thermometers responses. Note the frequency doubling, between the voltage excitation and the temperature evolution.

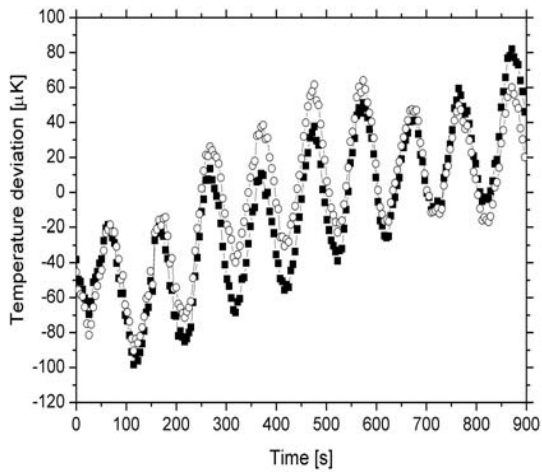

Fig. 4. Plots of two coherent responses of the observing thermometers. It shows a temperature oscillation of about 30 µK$_p$, superimposed to a temperature drift of about 100 nK/s.

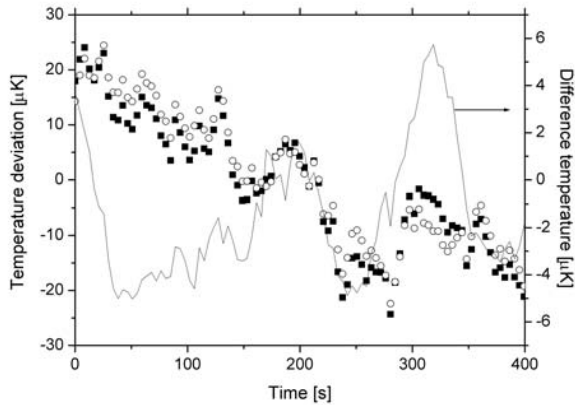

Fig. 5. Free response of the observing thermometers in the locked loop mode, referred by the left scale. The maximum difference between the thermometers reaches $\pm\ 6\ \mu K$ (right axis) with a standard deviation of 2.8 $\mu K_{rms}$.

## V. Conclusion

We have focused our work on the use of $YBa_2Cu_3O_{7-\delta}$ TES sensors. A standard deviation of 2.8 $\mu K_{rms}$ was achieved. The sensors can be use as well above their resistive transition, similar to platinum resistors. Although not characterized in this regime, we expect performances in the range of 25 to 30 $\mu K_{rms}$ as done for IPRT. The question of going below the temperature of the transition was also addressed a few years ago [6], but, without the pair of independent observing thermometers. Our knowing using such critical current sensors, is that it is much more difficult to design a low noise read-out electronics because of their low source impedance, however performances in the range of 1 to 10 $\mu K_{rms}$ should be feasible as well.

## Acknowledgment

We acknowledge Mr Poirier for carefully shaping the copper pieces of our sample holder, N. Chéenne for giving the first $YBa_2Cu_3O_{7-\delta}$ thermometers, and S. Lebargy for electronics.

## References

[1] J.P. Rice and Z. M. Zhang, "Liquid-Nitrogen-Cooled high-Tc electrical substitution as a broadband IR transfer standard", NIST Tech. Rep. 1414, 1996.
[2] J.P. Rice, S.R. Lorentz, R.U. Datla, L.R. Vale, D.A. Rudman, M. Lam Chok Sing and D. Robbes, "Active cavity absolute radiometer based on high Tc superconductors," *Metrologia*, 35(4), pp. 289-293, 1998.
[3] Gilbert Gaussorgues, in "*La thermographie infrarouge: principes technologies applications*," 4th ed. Ed. New York, Londres Paris: TEC&DOC, pp. 570-580, 1999.
[4] D. G. McDonald, R. J. Phelan, L. R. Vale, R. H. Ono and D. A. Rudman, "Passivation, Transition Width, and Noise for YBCO Bolometers on Silicon", *IEEE Trans. Appl. Supercond.,* vol. 9(2), pp. 4471-4474, June 1999.
[5] Circuit simulation logiciel (see for details http://www.orcad.com/products/pspice)
[6] M. Lam Chok Sing, E. Lesquey, C. Dolabdjian and D. Robbes, "A high stability temperature controller based on a high Tc sensor," *IEEE Trans. Appl. Supercond.,* vol. 7(2), pp. 3087-3090, June 1997.